\documentclass[twocolumn,showpacs,preprintnumbers,amsmath,amssymb]{revtex4}

\usepackage{graphicx}
\usepackage{dcolumn}
\usepackage{bm}
\usepackage{epsfig}

\begin{document}


\title{Detecting genuine multipartite continuous-variable entanglement}
\author{Peter van Loock$^1$ and Akira Furusawa$^2$}
\affiliation{$^1$ Quantum Information Theory Group,
Zentrum f\"{u}r Moderne Optik,
Universit\"{a}t Erlangen-N\"{u}rnberg,
91058 Erlangen, Germany\\
$^2$ Department of Applied Physics,
University of Tokyo, Tokyo 113-8656, Japan}

\begin{abstract}
We derive necessary conditions in terms of the variances of
position and momentum linear combinations
for all kinds of separability of a
multi-party multi-mode continuous-variable state.
Their violations can be sufficient for genuine multipartite
entanglement, provided the combinations contain both conjugate
variables of all modes.
Hence a complete state determination,
for example by detecting the entire correlation matrix of a Gaussian
state, is not needed.
\end{abstract}
\maketitle


\section{Introduction}

Quantum entanglement shared by two parties enhances their capability
to communicate. In principle, it allows them to convey quantum information
reliably through a classical channel (quantum teleportation \cite{Benn93}),
to double the amount of classical information transmittable through a classical channel
(quantum dense coding \cite{Wiesner92}), or to prepare at a distance states from nonorthogonal
bases for secure communication (quantum key distribution 
\cite{BB84,Ekert,BennMermin}).
These entanglement-assisted communication schemes are extendible to an arbitrary
number of parties sharing multipartite entanglement.
For instance, a sender may transfer quantum information through classical channels
to several receivers as reliably as allowed by optimal cloning
(telecloning \cite{Murao99}), or the parties may share quantum (or classical) information
retrievable only when all parties cooperate 
(quantum secret sharing \cite{Hillqsecret99}).
A more recent proposal that exploits the multi-party quantum correlations of
multipartite entangled states is the so-called Byzantine agreement protocol
\cite{Fitzi01}.
In general, the presence of entanglement is verified
through the success of a quantum protocol that would fail otherwise
(e.g., quantum teleportation).
Such an operational criterion is only sufficient for
entanglement and failure does not necessarily rule out its presence.
In an experimental realization, however, before running through an entire
entanglement-based protocol, it is desirable first to confirm that the generation
of sophisticated multi-party entangled states has succeeded.
The aim of this paper is to provide a simple but unambiguous experimental test
to check for a particular kind of genuinely multipartite entangled states, namely
those described by continuous variables (cv) and produced with squeezed light and
linear optics.

Work in the direction of generating
tripartite cv entanglement has been carried out
already by sending one half of a two-mode two-party entangled state 
through an extra beam splitter with a coherent state or a vacuum 
state at its second input port \cite{Furusawa98,Jing02}.
The resulting three-mode state was a side product of the
Bell measurement for the teleportation of coherent states
using a preshared symmetric two-mode squeezed state
\cite{Furusawa98}.
Its tripartite entanglement was not further investigated in that
experiment.
In another experiment, reported recently \cite{Jing02},
the two-mode state was asymmetric, corresponding to
two independently squeezed states combined at an asymmetric $1:2$
beam splitter. The output three-mode state after an additional 
symmetric beam splitter was then similar to the states proposed
in Ref.~\cite{PvLPRL00}.

Quantum communication, or more general quantum information with cv
has attracted a lot of interest due to the relative simplicity and high efficiency
in the generation, manipulation, and detection of optical cv
states. Although recent results suggest that these assets of Gaussian cv
operations (phase shifting, beam splitting, homodyne detections, phase-space
displacements, squeezing) are not extendible to more advanced quantum protocols
such as entanglement distillation \cite{Eisertdist,Fiurasekdist,Giedkedist}, 
the simple and efficient cv
approach still seems promising for many tasks and might be suitable for others too
when combined with discrete-variable (dv) strategies. On the other hand,
potential linear-optics implementations of quantum protocols solely based on dv
utilizing single photons are restricted by No-Go results such as the
impossibility of a complete distinction between the four 
Bell states \cite{Luetkenhaus99}.
In order to perform such a Bell measurement
near perfectly with linear optics,
one has to employ complicated entangled states of many auxiliary 
photons \cite{Knill01}.
In contrast, a Bell and also a GHZ state analyzer can be easily 
constructed in the
cv setting using only beam splitters and homodyne detectors
\cite{SamKimble98,vanloockFdP02,vanloockcvbook02}.

How may one now verify experimentally the presence of entanglement without
implementing a full quantum protocol?
We are here particularly concerned about the experimental verification
of genuinely multipartite entangled states where none of the parties
is separable from the rest (in terms of the separability properties of the
total density matrix).
In general, theoretical tests might be as well applicable to the experimental
verification. For instance, the violation of inequalities imposed by local realism
confirms the presence of entanglement.
Proving genuine multipartite entanglement,
however, requires stronger violations \cite{Seevinck} 
than those determined by the
commonly used Mermin-Klyshko $N$-party inequalities 
\cite{Mermin90,KlyshkoGisin}.
Moreover, in any case, violations
of Bell-type inequalities using Gaussian cv entangled states
with always positive Wigner functions must rely on observables other than the
quadratures (i.e., position and momentum). Photon number parity may serve
as an appropriate dichotomic variable to reveal the nonlocality of the
cv entangled states \cite{Bana98}. This applies to
the two-party two-mode EPR-like \cite{Bana98} and to the $N$-party $N$-mode
GHZ-like cv states \cite{PvLnonlocal01}. Such an approach, however,
is not very feasible due to its need for detectors resolving large photon number.

The negative partial transpose (npt) criterion \cite{Peres96} is
sufficient and necessary for the bipartite inseparability of
$2\times 2$-dimensional, $2\times 3$-dimensional \cite{Horodecki96},
and $1\times N$ mode Gaussian states \cite{Simon00,WernerWolf01}.
A complete experimental determination of the state in question
would also enable an npt check.
In general, any theoretical test is applicable when the experimentalist has
full information about the quantum state after measurements on an ensemble
of identically prepared states (e.g. by quantum tomography 
\cite{Leonhardtbook97}).
Such a {\it direct} verification of entanglement via a complete state
measurement is in general very demanding to the experimentalist,
in particular when the state to be determined is a potentially multi-party
entangled multi-mode state.

\section{Gaussian states}

The multi-party entanglement criteria 
that we will derive here do not rely on the assumption
of Gaussian states. However, the states commonly produced
in the laboratory are indeed Gaussian and the theoretical
classification of different types of multipartite entanglement
becomes simpler for Gaussian states \cite{Giedke01}.
 
Since the entanglement properties of a multi-mode multi-party
state are invariant under local phase-space displacements,
the multi-mode states may have zero mean and their Wigner 
function is of the form
\begin{eqnarray}\label{Gausswigndef}
W(\xi)=
\frac{1}{(2\pi)^N\sqrt{\det V^{(N)}}}\,\exp\left\{-\frac{1}{2}\,
\xi\left[V^{(N)}\right]^{-1}
\xi^{T}\right\}\;,\nonumber\\
\end{eqnarray}
with the $2N$-dimensional vector $\xi$ having
the quadrature pairs of all $N$ modes as its components,
\begin{eqnarray}
\xi&=&(x_1,p_1,x_2,p_2,...,x_N,p_N)\;,\nonumber\\
\hat\xi&=&(\hat x_1,\hat p_1,\hat x_2,\hat p_2,...,
\hat x_N,\hat p_N)\;,
\end{eqnarray}
and with the $2N\times 2N$ correlation matrix $V^{(N)}$ having
as its elements the second moments symmetrized according to 
the Weyl correspondence \cite{Weyl50},
\begin{eqnarray}\label{corrdef}
{\rm Tr}[\hat\rho\,(\Delta\hat\xi_i\Delta\hat\xi_j+
\Delta\hat\xi_j\Delta\hat\xi_i)/2]
&=&\langle(\hat\xi_i\hat\xi_j+\hat\xi_j\hat\xi_i)/2\rangle\nonumber\\
&=&\int\,W(\xi)\,\xi_i \xi_j\, d^{2N}\xi\nonumber\\
&=&V^{(N)}_{ij}\;,
\end{eqnarray}
where $\Delta\hat\xi_i=\hat\xi_i-\langle\hat\xi_i\rangle=\hat\xi_i$ for
zero mean values.
Note that the correlation matrix of any physical state
must be real, symmetric, positive, and must obey
the commutation relation \cite{Simon00,WernerWolf01},
\begin{eqnarray}
[\hat\xi_k,\hat\xi_l]=\frac{i}{2}\,\Lambda_{kl}\;,
\quad\quad k,l=1,2,3,...,2N\;,
\end{eqnarray}
with the $2N\times 2N$ matrix $\Lambda$ having
the $2\times 2$ matrix $J$ as diagonal entry for each
quadrature pair, for example for $N=2$, 
\begin{eqnarray}
\Lambda=
\left( \begin{array}{cc} J & 0 \\ 
0 & J 
\end{array} \right)\;,\quad\quad
J=
\left( \begin{array}{cc} 0 & 1 \\ 
-1 & 0 
\end{array} \right)\;. 
\end{eqnarray}
A direct consequence of this commutation relation and the non-negativity
of the density operator $\hat\rho$ is the $N$-mode
uncertainty relation \cite{Simon00,WernerWolf01},
\begin{eqnarray}\label{Nmodeuncert}
V^{(N)}-\frac{i}{4}\,\Lambda\geq 0\;.
\end{eqnarray}
Note that this condition is equivalent to 
$V^{(N)}+i\Lambda/4\geq 0$ by complex conjugation.
As for the direct verification of entanglement via a complete state
measurement, for Gaussian cv states, 
the complete measurement of an $N$-party $N$-mode
quantum state is accomplished by determining the $2N\times 2N$
second-moment correlation matrix.
This corresponds to $N(1+2N)$ independent entries taking into
account the symmetry of the correlation matrix.
Kim et al. \cite{Kim02} recently demonstrated how to determine all these
entries in the two-party two-mode case using beam splitters and homodyne detectors.
Joint homodyne detections of the two modes yield the intermode correlations
such as $\langle\hat x_1\hat x_2\rangle - \langle\hat x_1\rangle\langle\hat x_2\rangle$,
$\langle\hat x_1\hat p_2\rangle - \langle\hat x_1\rangle\langle\hat p_2\rangle$, etc.
Determining the local intramode correlations such as
$\langle\hat x_1\hat p_1+\hat p_1\hat x_1\rangle/2
- \langle\hat x_1\rangle\langle\hat p_1\rangle$
is more subtle and requires
additional beam splitters and homodyne detections
(or, alternatively, heterodyne detections).
Once the $4\times 4$ two-mode correlation matrix is known, the npt criterion
can be applied as a sufficient and necessary condition for bipartite Gaussian
two-mode inseparability (where npt corresponds to a sign change
of all the momentum variables with positions unchanged \cite{Simon00}).
In fact, the entanglement can also be quantified for a given correlation matrix
\cite{Kim02,VidalWerner02}.
For three-party three-mode Gaussian states, one may pursue a similar strategy.
After measuring the 21 independent entries of the correlation matrix
(for example, by extending Kim et al.'s scheme \cite{Kim02} to the three-mode case),
the sufficient and necessary criteria by Giedke et al. \cite{Giedke01} can be
applied. Let us examine the separability properties of 
(in particular, three-party three-mode) Gaussian states in more detail.

\section{Separability properties of Gaussian states}

The criteria by Giedke et al. \cite{Giedke01}
determine to which of five possible
classes of fully and partially separable, and fully inseparable states
a three-party three-mode Gaussian state belongs. 
Hence genuine tripartite entanglement if present
can be unambiguously identified.
The classification is mainly based on the 
npt criterion for cv states. Transposition  
is a positive map that corresponds in phase space
to a sign change of all momentum
variables, $\xi^{T}\rightarrow \Gamma\xi^{T}=
(x_1,-p_1,x_2,-p_2,...,x_N,-p_N)^{T}$ \cite{Simon00}.
In terms of the correlation matrix, we have then
$V^{(N)}\rightarrow\Gamma V^{(N)}\Gamma$.
Since transposition is not a completely positive map,
its partial application to a subsystem only
may yield an unphysical state when the subsystem
was entangled to other subsystems.
Expressing partial transposition of a bipartite 
Gaussian system by $\Gamma_a\equiv\Gamma\oplus
\mbox{1$\!\!${\large 1}}$ (where $A\oplus B$
means the block-diagonal matrix with the matrices
$A$ and $B$ as diagonal `entries', and $A$ and $B$
are respectively $2N\times 2N$ and $2M\times 2M$ square 
matrices applicable to $N$ modes at $a$'s side
and $M$ modes at $b$'s side),
the condition that the partially transposed
Gaussian state described by $\Gamma_a V^{(N+M)}\Gamma_a$
is unphysical [see Eq.~(\ref{Nmodeuncert})], 
$\Gamma_a V^{(N+M)}\Gamma_a\ngeq \frac{i}{4}\,\Lambda$,
is sufficient for the inseparability
between $a$ and $b$ \cite{Simon00,WernerWolf01}.
For Gaussian states with $N=1$ and arbitrary $M$, 
this condition is sufficient and necessary \cite{WernerWolf01}.
The simplest example where the condition is no longer
necessary for inseparability involves two modes at each
side, $N=M=2$. In that case, states with positive partial
transpose (bound entangled Gaussian states) exist
\cite{WernerWolf01}.
For the general bipartite $N\times M$ case,
there is also a sufficient and necessary condition:
the correlation matrix $V^{(N+M)}$ corresponds to
a separable state iff a pair of correlation matrices
$V^{(N)}_a$ and $V^{(M)}_b$ exists such that
$V^{(N+M)}\geq V^{(N)}_a\oplus V^{(M)}_b$.
Since it is in general hard to find such a pair
of correlation matrices $V^{(N)}_a$ and $V^{(M)}_b$ 
for a separable state or to prove the non-existence
of such a pair for an inseparable state, this criterion
in not very practical.
A more practical solution was provided in 
Ref.~\cite{Giedke01PRL}. The operational
criteria there, computable and testable via
a finite number of iterations,
are entirely independent of the npt criterion.
They rely on a nonlinear map between the correlation matrices
rather than a linear one such as the partial transposition,
and in contrast to the npt criterion,
they witness also the inseparability of bound entangled 
states.
Thus, the separability
problem for bipartite Gaussian states with arbitrarily many 
modes at each side is completely solved.
For three-party three-mode Gaussian states,
the only partially separable forms are those
with a bipartite splitting of $1 \times 2$ modes.
Hence already the npt criterion is sufficient and necessary.

The classification of tripartite three-mode Gaussian states
\cite{Giedke01},
\begin{eqnarray}
&&\quad{\rm class}\;1\,:\quad\quad
\bar V^{(3)}_1\ngeq \frac{i}{4}\,\Lambda\,,
\bar V^{(3)}_2\ngeq \frac{i}{4}\,\Lambda\,,
\bar V^{(3)}_3\ngeq \frac{i}{4}\,\Lambda\,,
\nonumber\\
&&\quad{\rm class}\;2\,:\quad\quad
\bar V^{(3)}_k\geq \frac{i}{4}\,\Lambda\,,
\bar V^{(3)}_m\ngeq \frac{i}{4}\,\Lambda\,,
\bar V^{(3)}_n\ngeq \frac{i}{4}\,\Lambda\,,
\nonumber\\
&&\quad{\rm class}\;3\,:\quad\quad
\bar V^{(3)}_k\geq \frac{i}{4}\,\Lambda\,,
\bar V^{(3)}_m\geq \frac{i}{4}\,\Lambda\,,
\bar V^{(3)}_n\ngeq \frac{i}{4}\,\Lambda\,,
\nonumber\\
&&{\rm class}\;4\;{\rm or}\;5\,:\quad
\bar V^{(3)}_1\geq \frac{i}{4}\,\Lambda\,,
\bar V^{(3)}_2\geq \frac{i}{4}\,\Lambda\,,
\bar V^{(3)}_3\geq \frac{i}{4}\,\Lambda\,,
\nonumber\\
\end{eqnarray} 
is solely based on the npt criterion, where
$\bar V^{(3)}_j\equiv\Gamma_j V^{(3)}\Gamma_j$ denotes
the partial transposition with respect to one mode $j$.
In classes 2 and 3, any permutation of modes ($k,m,n$)
must be considered.
Class 1 corresponds to the fully inseparable states.
Class 5 shall contain the fully separable states.
For the fully separable Gaussian states if described
by $V^{(3)}$, one-mode correlation matrices
$V^{(1)}_1$, $V^{(1)}_2$, and $V^{(1)}_3$ exist such that
$V^{(3)}\geq V^{(1)}_1\oplus V^{(1)}_2\oplus V^{(1)}_3$.
In general, fully separable quantum states can be written as
a mixture of tripartite product states, 
$\sum_i \eta_i\, \hat\rho_{i,1}\otimes\hat\rho_{i,2}
\otimes\hat\rho_{i,3}$.
In class 2, we have the one-mode biseparable states,
where only one particular mode is separable from
the remaining pair of modes.
This means in the Gaussian case that
only for one particular mode $k$,
$V^{(3)}\geq V^{(1)}_k\oplus V^{(2)}_{mn}$
with some two-mode correlation matrix $V^{(2)}_{mn}$
and one-mode correlation matrix $V^{(1)}_k$.
In general, such a state can be written as
$\sum_i \eta_i\, \hat\rho_{i,k}\otimes\hat\rho_{i,mn}$
for one mode $k$.
Class 3 contains those states where two but not three
bipartite splittings are possible, i.e.,
two different modes $k$ and $m$
are separable from the remaining 
pair of modes (two-mode biseparable states).
The states of class 4 (three-mode biseparable states)
can be written as a mixture of products
between any mode 1, 2, or 3
and the remaining pair of modes,
but not as a mixture of three-mode product states.
Obviously, classes 4 and 5 are not distinguishable
via the npt criterion.
An additional criterion for this distinction
of class 4 and 5 Gaussian states
is given in Ref.~\cite{Giedke01},
deciding whether one-mode correlation matrices
$V^{(1)}_1$, $V^{(1)}_2$, and $V^{(1)}_3$ exist such that
$V^{(3)}\geq V^{(1)}_1\oplus V^{(1)}_2\oplus V^{(1)}_3$.
For the identification of genuinely tripartite entangled
Gaussian states, only class 1 has to be distinguished
from the rest. Hence the npt criterion alone suffices.

What about more than three parties and modes?
Even for only four parties and modes, the separability
issue becomes more subtle.
The one-mode bipartite splittings,
$\sum_i \eta_i\, \hat\rho_{i,klm}\otimes\hat\rho_{i,n}$,
can be tested and possibly ruled out via the npt criterion
with respect to any mode $n$.
In the Gaussian language,
if $\bar V^{(4)}_n\ngeq\frac{i}{4}\,\Lambda$
for any $n$, the state cannot be written in the above form.
Since we consider here the bipartite splitting of 
$1\times 3$ modes, the npt condition is sufficient and necessary
for Gaussian states.
However, also a state of the form
$\sum_i \eta_i\, \hat\rho_{i,kl}\otimes\hat\rho_{i,mn}$
leads to negative partial transpose with respect to any of the
four modes when the two pairs ($k,l$) and ($m,n$) are each
entangled. Thus, npt with respect to any individual mode
is necessary but not sufficient for genuine four-party entanglement.
One has to consider also the partial transposition
with respect to any pair of modes.
For this $2\times 2$ mode case, however, we know that
entangled Gaussian states with positive partial transpose exist
\cite{WernerWolf01}. But the npt criterion is still sufficient
for the inseparability between any two pairs.
As for a sufficient and necessary condition, one can use
those from Ref.~\cite{Giedke01PRL}.
In any case, in order to confirm genuine four-party
or even $N$-party entanglement, one has to rule out
any possible partially separable form. 
In principle, this can be done by considering all possible
bipartite splittings (or groupings)
and applying either the npt criterion
or the stronger operational criteria from Ref.~\cite{Giedke01PRL}.
Although a full theoretical characterization including criteria for 
entanglement classification has not been considered yet for more than 
three parties and modes,
the presence of genuine multipartite entanglement can be confirmed,
once the complete $2N\times 2N$ correlation matrix is given.

Rather than detecting all the entries of the correlation matrix
we are aiming here at a simple check
based on only a few measurements, preferably efficient homodyne detections.
Even for larger numbers of parties, this check should remain simple.
Though it may not yield full information (e.g., the complete correlation
matrix) about the quantum state of interest, it should still unambiguously
verify the presence of genuine multipartite entanglement.
This check may prove the presence of entanglement {\it indirectly}
through measurements after transforming the relevant state first into
an appropriate form via linear optics.

\section{Detecting entanglement: bipartite case}

In the two-party two-mode case, the necessary separability
condition for any cv state \cite{Duan00}
\begin{equation}\label{2partycrit}
\langle[\Delta(\hat{x}_1-\hat{x}_2)]^2\rangle +
\langle[\Delta(\hat{p}_1+\hat{p}_2)]^2\rangle \geq 2\,|\langle[\hat x,\hat p]
\rangle | \;,
\end{equation}
can be tested, for example, with a single beam splitter.
The position and momentum variables $\hat x_l$ and $\hat p_l$
(units-free with $\hbar=\frac{1}{2}$,
$[\hat{x}_l,\hat{p}_k]=i\delta_{lk}/2$)
correspond to the quadratures of two electromagnetic
modes, i.e., the real and imaginary parts of the
annihilation operators of the two modes:
$\hat{a}_l=\hat{x}_l+i\hat{p}_l$.
The beam splitter provides the suitable
quadrature combinations for the positions and momenta simultaneously
detectable at the two output ports. Without beam splitter,
just by measuring first both positions and subtracting them electronically,
and in a second step detecting both momenta and combining these
electronically \cite{akiracvbook02}, a more direct test of the two-party
condition is possible. However, instead of a simultaneous detection
of the relevant combinations, it requires switching the two local oscillator
phases from position to momentum measurements.
For an ensemble of identically prepared states, this sequence of detections
would still enable the application of the two-party condition.
Note that the violation of Eq.~(\ref{2partycrit}) is only sufficient for
inseparability, i.e., there are (even Gaussian) cv entangled states
that satisfy Eq.~(\ref{2partycrit}). Any Gaussian cv state, however,
can be transformed via local operations into a standard form and the
presence of entanglement would then always yield a violation \cite{Duan00}
(alternatively, one may modify the inequality and leave the Gaussian state
unchanged to obtain a sufficient and necessary condition \cite{Giovannetti02}).
The point is that the entanglement of states already
in this standard form (such as two-mode squeezed states)
can, in principle, always (for any nonzero squeezing) be verified
experimentally by checking Eq.~(\ref{2partycrit}).
A full determination of the correlation matrix,
including elements such as
$\langle\hat x_1\hat p_2\rangle - \langle\hat x_1\rangle\langle\hat p_2\rangle$
which do not appear in the expressions of Eq.~(\ref{2partycrit}),
is not required.
Measuring also these elements may confirm that the
state is in standard form (when they are zero) and hence
render the condition Eq.~(\ref{2partycrit}) sufficient and necessary
for separability. In any case, it would also enable quantification
of the entanglement \cite{Kim02,VidalWerner02}.

The combinations in condition Eq.~(\ref{2partycrit}) are exactly
those detected in a cv Bell measurement of modes 1 and 2 \cite{SamKimble98}.
Thus, the verification of non-maximum two-mode cv entanglement
may rely on measurements of observables that are detected for
the projection onto the maximally entangled cv basis of two modes.
Now we investigate the $N$-party $N$-mode case
in that respect.

\section{The cv GHZ basis}
\label{cvGHZ}

Let us introduce the maximally entangled states
\begin{eqnarray}
|\Psi(v,u_1,u_2,...,u_{N-1})\rangle =
\frac{1}{\sqrt{\pi}}\int_{-\infty}^{\infty}\, dx\, e^{2ivx}
\nonumber\\
\label{PVLmaxentGHZbasis}
\times |x\rangle\otimes|x-u_1\rangle
\otimes\,|x-u_1-u_2\rangle \quad\;\;\;\;\,
\nonumber\\
\otimes\!\!\cdots\!\!\otimes\,
|x-u_1-u_2-\cdots-u_{N-1}
\rangle\;.
\end{eqnarray}
Since $\int_{-\infty}^{\infty}\,|x\rangle\langle x|
=\mbox{1$\!\!${\large 1}}$ and
$\langle x|x'\rangle=\delta(x-x')$,
they form a complete,
\begin{eqnarray}
&&\int_{-\infty}^{\infty}\,dv\,du_1\,du_2\cdots du_{N-1}
\\
&&\times|\Psi(v,u_1,u_2,...,u_{N-1})\rangle\langle
\Psi(v,u_1,u_2,...,u_{N-1})|
=\mbox{1$\!\!${\large 1}}^{\otimes N},\nonumber
\end{eqnarray}
and orthogonal,
\begin{eqnarray}
&&\!\!\!\!\!\!\!\!\!
\langle\Psi(v,u_1,u_2,...,u_{N-1})|
\Psi(v',u_1',u_2',...,u_{N-1}')\rangle
\\
&&=\delta(v-v')\delta(u_1-u_1')\delta(u_2-u_2')
\cdots\delta(u_{N-1}-u_{N-1}'),\nonumber
\end{eqnarray}
set of basis states for $N$ modes.
In a ``cv GHZ state analyzer'',
determining the quantities $v\equiv p_1+p_2+\cdots+p_N$,
$u_1\equiv x_1-x_2$, $u_2\equiv x_2-x_3$,...,
and $u_{N-1}\equiv x_{N-1}-x_N$ means
projecting onto the
basis $\{|\Psi(v,u_1,u_2,...,u_{N-1})\rangle\}$.
This can be accomplish with a sequence of beam splitters
and homodyne detections \cite{vanloockFdP02,vanloockcvbook02}.
Inferring from the two-party case, we may conjecture that
the $N$ quadrature combinations given by
$v,u_1,u_2,...,u_{N-1}$ provide a sufficient set of observables
for the verification of (possibly genuine) $N$-party entanglement.
Just as for two parties, the variances of these quantities could
then also be determined by combining the results of
direct $x$ and $p$ measurements electronically.
It was shown in Ref.~\cite{vanloockFdP02,vanloockcvbook02} that
conditions for genuine multipartite entanglement can be derived
based on the above $N$ combinations and additional assumptions
such as the purity and the total symmetry of the state in question.
Later we derive a set of $N-1$ conditions for those $N$ combinations
sufficient for the presence of genuine multipartite
entanglement. This set is well suited for the experimental
confirmation of the genuine multi-party entanglement
of cv GHZ-type states. No extra assumptions
about the state are needed in order to close the loophole of
partial separability.
First, we discuss now what the structure of 
simple experimental criteria
for multipartite cv entanglement might be.

\section{Detecting entanglement: tripartite case}

Let us consider three parties and modes.
The goal is to extend the simple two-party two-mode entanglement check
to a simple test for genuine three-party three-mode entanglement.
The criteria are to be expressed in terms of the variances
of quadrature linear combinations for the modes involved.
Defining
\begin{eqnarray}\label{threemodecombin}
\hat u\equiv h_1\hat x_1 + h_2\hat x_2 + h_3\hat x_3 \;,
\hat v\equiv g_1\hat p_1 + g_2\hat p_2 + g_3\hat p_3 \;,
\end{eqnarray}
a fairly general ansatz is
\begin{eqnarray}\label{3partycritgenansatz}
\langle(\Delta\hat{u})^2\rangle_{\rho}+
\langle(\Delta\hat{v})^2\rangle_{\rho}\geq
f(h_1,h_2,h_3,g_1,g_2,g_3)\;,
\end{eqnarray}
as a potential necessary condition for an at least partially
separable state.
The position and momentum variables $\hat x_l$ and $\hat p_l$
are the quadratures of the three electromagnetic modes.
The $h_l$ and $g_l$ are arbitrary real parameters.
We will prove the following statements
for (at least partially) separable states,
\begin{eqnarray}\label{3partystatements}
\hat\rho&=&\sum_i \eta_i\, \hat\rho_{i,12}\otimes\hat\rho_{i,3}
\nonumber\\
\label{statem1}
&& \rightarrow \;
f(h_l,g_l)=(|h_3 g_3| + |h_1 g_1 + h_2 g_2|)/2 \,,
\\
\hat\rho&=&\sum_i \eta_i\, \hat\rho_{i,13}\otimes\hat\rho_{i,2}
\nonumber\\
\label{statem2}
&& \rightarrow \;
f(h_l,g_l)=(|h_2 g_2| + |h_1 g_1 + h_3 g_3|)/2 \,,
\\
\hat\rho&=&\sum_i \eta_i\, \hat\rho_{i,23}\otimes\hat\rho_{i,1}
\nonumber\\
\label{statem3}
&& \rightarrow \;
f(h_l,g_l)=(|h_1 g_1| + |h_2 g_2 + h_3 g_3|)/2 \,.
\end{eqnarray}
Here, for instance, $\hat\rho_{i,12}\otimes\hat\rho_{i,3}$ indicates
that the three-party density operator is a mixture of states $i$
where parties (modes) 1 and 2 may be entangled or not, but party 3
is not entangled with the rest. Hence also the fully separable state
is included in the above statements.
In fact, for the fully separable state, we have
\begin{eqnarray}\label{3partyfullysep}
\hat\rho&=&\sum_i \eta_i\, \hat\rho_{i,1}\otimes\hat\rho_{i,2}
\otimes\hat\rho_{i,3}
\nonumber\\
&& \rightarrow \;
f(h_l,g_l)=(|h_1 g_1| + |h_2 g_2| + |h_3 g_3|)/2,
\end{eqnarray}
which is always greater or equal than any of the
boundaries in Eq.~(\ref{statem1}),
Eq.~(\ref{statem2}), or Eq.~(\ref{statem3}).
For the proof, let us assume that the relevant state
can be written as
\begin{eqnarray}\label{3partyassumption}
\hat\rho=\sum_i \eta_i\, \hat\rho_{i,km}\otimes\hat\rho_{i,n}\;.
\end{eqnarray}
For the combinations in Eq.~(\ref{threemodecombin}), we find
\begin{eqnarray}\label{derivation1}
&&\langle(\Delta\hat{u})^2\rangle_{\rho}+
\langle(\Delta\hat{v})^2\rangle_{\rho}
\nonumber\\
&=&\sum_i \eta_i\; \left(\langle\hat{u}^2\rangle_i+
\langle\hat{v}^2\rangle_i\right)-
\langle\hat{u}\rangle_{\rho}^2-\langle\hat{v}\rangle_{\rho}^2
\nonumber\\
&=&\sum_i \eta_i\;
\Big[h_k^2\langle\hat{x}_k^2\rangle_i+
h_m^2\langle\hat{x}_m^2\rangle_i+
h_n^2\langle\hat{x}_n^2\rangle_i
\nonumber\\
&&\quad\quad\quad+g_k^2\langle\hat{p}_k^2\rangle_i+
g_m^2\langle\hat{p}_m^2\rangle_i+
g_n^2\langle\hat{p}_n^2\rangle_i
\nonumber\\
&&+2\Big(
h_k h_m\langle\hat{x}_k\hat{x}_m\rangle_i+
h_k h_n\langle\hat{x}_k\hat{x}_n\rangle_i+
h_m h_n\langle\hat{x}_m\hat{x}_n\rangle_i\Big)
\nonumber\\
&&+2\Big(
g_k g_m\langle\hat{p}_k\hat{p}_m\rangle_i+
g_k g_n\langle\hat{p}_k\hat{p}_n\rangle_i+
g_m g_n\langle\hat{p}_m\hat{p}_n\rangle_i\Big)\Big]
\nonumber\\
&&-\langle\hat{u}\rangle_{\rho}^2-\langle\hat{v}\rangle_{\rho}^2
\nonumber\\
&=&\sum_i \eta_i\;
\Big[
h_k^2\langle(\Delta\hat{x}_k)^2\rangle_i+
h_m^2\langle(\Delta\hat{x}_m)^2\rangle_i+
h_n^2\langle(\Delta\hat{x}_n)^2\rangle_i
\nonumber\\
&&\quad\quad\quad+
g_k^2\langle(\Delta\hat{p}_k)^2\rangle_i+
g_m^2\langle(\Delta\hat{p}_m)^2\rangle_i+
g_n^2\langle(\Delta\hat{p}_n)^2\rangle_i
\nonumber\\
&&\quad\quad\quad+2h_k h_m\Big(
\langle\hat{x}_k\hat{x}_m\rangle_i-
\langle\hat{x}_k\rangle_i\langle\hat{x}_m\rangle_i\Big)
\nonumber\\
&&\quad\quad\quad+2h_k h_n\Big(
\langle\hat{x}_k\hat{x}_n\rangle_i-
\langle\hat{x}_k\rangle_i\langle\hat{x}_n\rangle_i\Big)
\nonumber\\
&&\quad\quad\quad+2h_m h_n\Big(
\langle\hat{x}_m\hat{x}_n\rangle_i-
\langle\hat{x}_m\rangle_i\langle\hat{x}_n\rangle_i\Big)
\nonumber\\
&&\quad\quad\quad+2g_k g_m\Big(
\langle\hat{p}_k\hat{p}_m\rangle_i-
\langle\hat{p}_k\rangle_i\langle\hat{p}_m\rangle_i\Big)
\nonumber\\
&&\quad\quad\quad+2g_k g_n\Big(
\langle\hat{p}_k\hat{p}_m\rangle_i-
\langle\hat{p}_k\rangle_i\langle\hat{p}_m\rangle_i\Big)
\nonumber\\
&&\quad\quad\quad+2g_m g_n\Big(
\langle\hat{p}_m\hat{p}_n\rangle_i-
\langle\hat{p}_m\rangle_i\langle\hat{p}_n\rangle_i\Big)
\Big]
\nonumber\\
&&+\sum_i \eta_i\; \langle\hat{u}\rangle_i^2-\left(\sum_i \eta_i\;
\langle\hat{u}\rangle_i\right)^2
\nonumber\\
&&+\sum_i \eta_i\; \langle\hat{v}
\rangle_i^2-\left(\sum_i \eta_i\;
\langle\hat{v}\rangle_i\right)^2\,,
\end{eqnarray}
where $\langle\cdots\rangle_i$ means the average in the state
$\hat\rho_{i,km}\otimes\hat\rho_{i,n}$.
Note that in the derivation so far we have not used 
the particular form in Eq.~(\ref{3partyassumption}) yet.
Exploiting this form of the state, we obtain 
$\langle\hat{x}_k\hat{x}_n\rangle_i=
\langle\hat{x}_k\rangle_i\langle\hat{x}_n\rangle_i$,
$\langle\hat{x}_m\hat{x}_n\rangle_i=
\langle\hat{x}_m\rangle_i\langle\hat{x}_n\rangle_i$,
and similarly for the terms involving $p$.
Because modes $k$ and $m$ may be entangled
in the states $i$, we cannot replace
$\langle\hat{x}_k\hat{x}_m\rangle_i$ by
$\langle\hat{x}_k\rangle_i\langle\hat{x}_m\rangle_i$, etc.
By applying the Cauchy-Schwarz inequality as in the
two-party derivation of Ref.~\cite{Duan00},
$\sum_i P_i \langle\hat{u}\rangle_i^2 \geq \left(\sum_i P_i
|\langle\hat{u}\rangle_i|\right)^2$, we see that the last two lines
in Eq.~(\ref{derivation1}) are bounded below by zero.
Hence in order to prove
$\langle(\Delta\hat{u})^2\rangle_{\rho}+
\langle(\Delta\hat{v})^2\rangle_{\rho}\geq 
(|h_n g_n| + |h_k g_k + h_m g_m|)/2$,
it remains to be shown that for any $i$
[recall that the mixture in Eq.~(\ref{3partyassumption})
is a convex sum with $\sum_i\eta_i=1$],
\begin{eqnarray}\label{derivation2}
&&h_k^2\langle(\Delta\hat{x}_k)^2\rangle_i+
h_m^2\langle(\Delta\hat{x}_m)^2\rangle_i+
h_n^2\langle(\Delta\hat{x}_n)^2\rangle_i
\nonumber\\
&&+
g_k^2\langle(\Delta\hat{p}_k)^2\rangle_i+
g_m^2\langle(\Delta\hat{p}_m)^2\rangle_i+
g_n^2\langle(\Delta\hat{p}_n)^2\rangle_i
\nonumber\\
&&+2h_k h_m\Big(
\langle\hat{x}_k\hat{x}_m\rangle_i-
\langle\hat{x}_k\rangle_i\langle\hat{x}_m\rangle_i\Big)
\nonumber\\
&&+2g_k g_m\Big(
\langle\hat{p}_k\hat{p}_m\rangle_i-
\langle\hat{p}_k\rangle_i\langle\hat{p}_m\rangle_i\Big)
\nonumber\\
&&\geq (|h_n g_n| + |h_k g_k + h_m g_m|)/2\;.
\end{eqnarray}
By rewriting the left-hand-side of Eq.~(\ref{derivation2})
in terms of variances only, indeed we find 
\begin{eqnarray}\label{derivation3}
&&h_n^2\langle(\Delta\hat{x}_n)^2\rangle_i+
g_n^2\langle(\Delta\hat{p}_n)^2\rangle_i
\nonumber\\
&&+\langle[\Delta(h_k\hat{x}_k + h_m\hat x_m)]^2\rangle_i
+\langle[\Delta(g_k\hat{p}_k + g_m \hat p_m)]^2\rangle_i
\nonumber\\
&&\geq
|\langle [h_n\hat x_n,g_n \hat p_n]\rangle |+
|\langle [h_k\hat x_k + h_m\hat x_m,g_k\hat{p}_k 
+ g_m\hat p_m]\rangle |
\nonumber\\
&&=(|h_n g_n| + |h_k g_k + h_m g_m|)/2\;,
\end{eqnarray}
using the sum uncertainty relation
$\langle(\Delta\hat{A})^2\rangle+\langle(\Delta\hat{B})^2\rangle
\geq |\langle [\hat{A},\hat{B}]\rangle |$
and $[\hat{x}_l,\hat{p}_j]=i\delta_{lj}/2$.
Hence the statements in Eq.~(\ref{3partystatements})
are proven when we consider the corresponding
permutations of $(k,m,n)=(1,2,3)$.
The inequalities Eq.~(\ref{3partycritgenansatz}) with
Eq.~(\ref{statem1}), Eq.~(\ref{statem2}), and 
Eq.~(\ref{statem3})
represent necessary conditions for all kinds
of (partial) separability in a tripartite three-mode
state. One may then prove the presence of genuine
tripartite entanglement through violations of these 
inequalities, thus ruling out any (partially) separable
form. Whether there are really three different conditions
required for the verification depends on the choice
of the coefficients $h_l$ and $g_l$ in the linear
combinations. For a particular choice, some of the
conditions may coincide.
For example, consider
$h_1=g_1=1$ and $g_2=g_3=-h_2=-h_3=1/\sqrt{2}$
in Eq.~(\ref{threemodecombin}).
In this case, the boundaries in 
Eq.~(\ref{statem1}) and Eq.~(\ref{statem2}) become
identical, $f(h_l,g_l)=1/2$.
The boundary of Eq.~(\ref{statem3}) is even larger,
$f(h_l,g_l)=1$, equivalent to that for a fully
separable state in Eq.~(\ref{3partyfullysep}).
Hence the violation of a {\it single} condition,
\begin{eqnarray}\label{threepartyexample}
&&\langle\{\Delta[\hat x_1-(\hat x_2+\hat x_3)/\sqrt{2}]\}^2
\rangle_{\rho}\nonumber\\
&&\;+\langle\{\Delta[\hat p_1+(\hat p_2+\hat p_3)/\sqrt{2}]\}^2
\rangle_{\rho}\geq 1/2\;,
\end{eqnarray}
is already sufficient for genuine tripartite entanglement.
These particular combinations are not only significant
for the reason that they yield nonzero boundaries
for all kinds of separable states.
Moreover, their commutator vanishes,
\begin{eqnarray}
[\hat x_1-(\hat x_2+\hat x_3)/\sqrt{2},
\hat p_1+(\hat p_2+\hat p_3)/\sqrt{2}]=0\;,
\end{eqnarray}
allowing for arbitrarily good violations of
Eq.~(\ref{threepartyexample})
and, in principle, the existence of a simultaneous eigenstate
of these two combinations.
Such a state corresponds to the three-mode state obtainable
by splitting one half of an infinitely squeezed two-mode squeezed
(EPR) state at a 50:50 beam splitter.
The EPR correlations, $\hat x_1 - \hat x_2\rightarrow 0$ and
$\hat p_1 + \hat p_2\rightarrow 0$, are then transformed into
the three-mode correlations
$\hat x_1 - (\hat x_2' + \hat x_3')/\sqrt{2}\rightarrow 0$ and
$\hat p_1 + (\hat p_2' + \hat p_3')/\sqrt{2}\rightarrow 0$.
Let us turn to an arbitrary number of parties (modes)
now.

\section{Detecting entanglement: multipartite case}

Inferring from the discussion of the previous section,
the recipe for verifying the genuine multipartite
entanglement between arbitrarily
many parties and modes is the following.
First, measure both quadratures $x$ and $p$ of all modes involved
and combine them in an appropriate linear combination.
The variances of these combinations may then yield violations
of conditions necessary for partial separability.
Appropriate combinations
are those where the total variances for all partially separable
states have nonzero lower bounds
and where the commutators of the combinations vanish.
As for the derivation of the corresponding entanglement
criteria, we employ the following steps.

1. Select a distinct pair of modes $(m,n)$.

2. Choose appropriate linear combinations of the quadratures
in order to rule out all possible separable splittings between this pair
of modes in the convex sum of the total density operator.

3. Consider different pairs $(m,n)$ to negate all partial separabilities;
if necessary add further conditions involving other linear combinations.

Below it will become clear that step 2 can be performed simply by using the
appropriate bipartite combinations, $\hat x_m - \hat x_n$ and
$\hat p_m + \hat p_n$, i.e., by taking all $h_l=g_l=0$ except $h_m=g_m=1$
and $h_n=-g_n=-1$ in the general combinations
\begin{eqnarray}\label{multimodecombin}
\hat u&\equiv& h_1\hat x_1 + h_2\hat x_2 +\cdots +
h_N\hat x_N \;,\nonumber\\
\hat v&\equiv& g_1\hat p_1 + g_2\hat p_2 + \cdots +
g_N\hat p_N \;.
\end{eqnarray}
The boundaries of the total variance conditions are then identical
for any pair $(m,n)$ separable in the convex sum,
namely $f(h_l,g_l)\equiv 1$ in
\begin{eqnarray}\label{multipartycritgenansatz}
\langle(\Delta\hat{u})^2\rangle_{\rho}+
\langle(\Delta\hat{v})^2\rangle_{\rho}\geq
f(h_1,h_2,...,h_N,g_1,g_2,...,g_N)\;.\nonumber\\
\end{eqnarray}
However, in general, one obtains better multi-party conditions
when linear combinations for the quadratures of more than only
two modes are used. Through such multi-mode combinations the potential
multi-mode correlations are taken into account.
Before giving an example, let us first derive the general $N$-party
bounds in the condition Eq.~(\ref{multipartycritgenansatz}).
For any partially separable form, the total density operator
can be written as
\begin{eqnarray}\label{Npartyassumption}
\hat\rho=\sum_i \eta_i\, \hat\rho_{i,k_r \cdots m}
\otimes\hat\rho_{i,k_s\cdots n}\;,
\end{eqnarray}
with a distinct pair of ``separable modes'' $(m,n)$ and
the other modes $k_r\neq k_s$.
For the combinations in Eq.~(\ref{multimodecombin}), we find now
\begin{eqnarray}\label{multiderivation1}
&&\langle(\Delta\hat{u})^2\rangle_{\rho}+
\langle(\Delta\hat{v})^2\rangle_{\rho}
\nonumber\\
&=&\sum_i \eta_i\; \left(\langle\hat{u}^2\rangle_i+
\langle\hat{v}^2\rangle_i\right)-
\langle\hat{u}\rangle_{\rho}^2-\langle\hat{v}\rangle_{\rho}^2
\nonumber\\
&=&\sum_i \eta_i\;\Big[
h_m^2\langle\hat{x}_m^2\rangle_i+
h_n^2\langle\hat{x}_n^2\rangle_i+
\sum_{j=1}^{N-2}h_{k_j} ^2
\langle\hat{x}_{k_j}^2\rangle_i
\nonumber\\
&&\quad\quad\quad+
g_m^2\langle\hat{p}_m^2\rangle_i+
g_n^2\langle\hat{p}_n^2\rangle_i+
\sum_{j=1}^{N-2}g_{k_j} ^2
\langle\hat{p}_{k_j}^2\rangle_i
\nonumber\\
&&+\sum_{j\neq j'=1}^{N-2}\Big(
h_{k_j} h_{k_{j'}}\langle\hat{x}_{k_j}\hat{x}_{k_{j'}}\rangle_i+
g_{k_j} g_{k_{j'}}\langle\hat{p}_{k_j}\hat{p}_{k_{j'}}\rangle_i
\Big)
\nonumber\\
&&+2\sum_{j=1}^{N-2}\Big(
h_{k_j} h_m\langle\hat{x}_{k_j}\hat{x}_m\rangle_i+
h_{k_j} h_n\langle\hat{x}_{k_j}\hat{x}_n\rangle_i
\nonumber\\
&&\quad\quad\quad\quad+
g_{k_j} g_m\langle\hat{p}_{k_j}\hat{p}_m\rangle_i+
g_{k_j} g_n\langle\hat{p}_{k_j}\hat{p}_n\rangle_i
\Big)
\nonumber\\
&&+2\Big(
h_m h_n\langle\hat{x}_m\hat{x}_n\rangle_i+
g_m g_n\langle\hat{p}_m\hat{p}_n\rangle_i\Big)\Big]
\nonumber\\
&&-\langle\hat{u}\rangle_{\rho}^2-\langle\hat{v}\rangle_{\rho}^2
\nonumber\\
&=&\sum_i \eta_i\;
\Big\{
h_m^2\langle(\Delta\hat{x}_m)^2\rangle_i+
h_n^2\langle(\Delta\hat{x}_n)^2\rangle_i
\nonumber\\
&&\quad\quad\quad+
g_m^2\langle(\Delta\hat{p}_m)^2\rangle_i+
g_n^2\langle(\Delta\hat{p}_n)^2\rangle_i
\nonumber\\
&&\quad\quad\quad+
\sum_{j=1}^{N-2}\Big(
h_{k_j}^2\langle(\Delta\hat{x}_{k_j})^2\rangle_i+
g_{k_j}^2\langle(\Delta\hat{p}_{k_j})^2\rangle_i
\Big)
\nonumber\\
&&\quad\quad\quad+
\sum_{r\neq r'}\Big[
h_{k_r} h_{k_{r'}}\Big(
\langle\hat{x}_{k_r}\hat{x}_{k_{r'}}\rangle_i-
\langle\hat{x}_{k_r}\rangle_i\langle\hat{x}_{k_{r'}}\rangle_i
\Big)
\nonumber\\
&&\quad\quad\quad\quad\quad\quad+
g_{k_r} g_{k_{r'}}\Big(
\langle\hat{p}_{k_r}\hat{p}_{k_{r'}}\rangle_i-
\langle\hat{p}_{k_r}\rangle_i\langle\hat{p}_{k_{r'}}\rangle_i
\Big)\Big]
\nonumber\\
&&\quad\quad\quad+
\sum_{s\neq s'}\Big[
h_{k_s} h_{k_{s'}}\Big(
\langle\hat{x}_{k_s}\hat{x}_{k_{s'}}\rangle_i-
\langle\hat{x}_{k_s}\rangle_i\langle\hat{x}_{k_{s'}}\rangle_i
\Big)
\nonumber\\
&&\quad\quad\quad\quad\quad\quad+
g_{k_s} g_{k_{s'}}\Big(
\langle\hat{p}_{k_s}\hat{p}_{k_{s'}}\rangle_i-
\langle\hat{p}_{k_s}\rangle_i\langle\hat{p}_{k_{s'}}\rangle_i
\Big)\Big]
\nonumber\\
&&\quad\quad\quad+
2\sum_{r}\Big[
h_{k_r} h_m\Big(
\langle\hat{x}_{k_r}\hat{x}_m\rangle_i-
\langle\hat{x}_{k_r}\rangle_i\langle\hat{x}_m\rangle_i
\Big)
\nonumber\\
&&\quad\quad\quad\quad\quad\quad+
g_{k_r} g_m\Big(
\langle\hat{p}_{k_r}\hat{p}_m\rangle_i-
\langle\hat{p}_{k_r}\rangle_i\langle\hat{p}_m\rangle_i
\Big)\Big]
\nonumber\\
&&\quad\quad\quad+
2\sum_{s}\Big[
h_{k_s} h_n\Big(
\langle\hat{x}_{k_s}\hat{x}_n\rangle_i-
\langle\hat{x}_{k_s}\rangle_i\langle\hat{x}_n\rangle_i
\Big)
\nonumber\\
&&\quad\quad\quad\quad\quad\quad+
g_{k_s} g_n\Big(
\langle\hat{p}_{k_s}\hat{p}_n\rangle_i-
\langle\hat{p}_{k_s}\rangle_i\langle\hat{p}_n\rangle_i
\Big)\Big]\Big\}
\nonumber\\
&&+\sum_i \eta_i\; \langle\hat{u}\rangle_i^2-\left(\sum_i \eta_i\;
\langle\hat{u}\rangle_i\right)^2
\nonumber\\
&&+\sum_i \eta_i\; \langle\hat{v}
\rangle_i^2-\left(\sum_i \eta_i\;
\langle\hat{v}\rangle_i\right)^2\,.
\end{eqnarray}
For the last equality, we exploited Eq.~(\ref{Npartyassumption}),
namely that modes $k_r$ through $m$ are separable from
modes $k_s$ through $n$ in the convex sum of the total density
operator.
Similar to the three-party case,
we can now apply the Cauchy-Schwarz inequality
to the last two lines of Eq.~(\ref{multiderivation1})
and express the remaining terms by variances only.
This leads for any $i$ to
\begin{eqnarray}\label{multiderivation2}
&&\Big\langle\Big[\Delta\Big(h_m\hat{x}_m +
\sum_r h_{k_r}\hat x_{k_r}\Big)\Big]^2\Big\rangle_i
\nonumber\\
&&+\Big\langle\Big[\Delta\Big(g_m\hat{p}_m +
\sum_r g_{k_r}\hat p_{k_r}\Big)\Big]^2\Big\rangle_i
\nonumber\\
&&+\Big\langle\Big[\Delta\Big(h_n\hat{x}_n +
\sum_s h_{k_s}\hat x_{k_s}\Big)\Big]^2\Big\rangle_i
\nonumber\\
&&+\Big\langle\Big[\Delta\Big(g_n\hat{p}_n +
\sum_s g_{k_s}\hat p_{k_s}\Big)\Big]^2\Big\rangle_i
\nonumber\\
&&\geq
\Big|\Big\langle \Big[h_m\hat x_m +
\sum_r h_{k_r}\hat x_{k_r},g_m \hat p_m +
\sum_r g_{k_r}\hat p_{k_r}\Big]\Big\rangle \Big|
\nonumber\\
&&\quad\,+\Big|\Big\langle \Big[h_n\hat x_n +
\sum_s h_{k_s}\hat x_{k_s},g_n \hat p_n +
\sum_s g_{k_s}\hat p_{k_s}\Big]\Big\rangle \Big|\;,
\nonumber\\
\end{eqnarray}
using again the sum uncertainty relation
$\langle(\Delta\hat{A})^2\rangle+\langle(\Delta\hat{B})^2\rangle
\geq |\langle [\hat{A},\hat{B}]\rangle |$.
Thus, by evaluating the commutators with
$[\hat{x}_l,\hat{p}_j]=i\delta_{lj}/2$,
we obtain for the total variance
\begin{eqnarray}\label{finalmultiparty}
&&\langle(\Delta\hat{u})^2\rangle_{\rho}+
\langle(\Delta\hat{v})^2\rangle_{\rho}
\nonumber\\
&&\quad\geq\frac{1}{2}\,
\Big(\Big| h_m g_m +
\sum_r h_{k_r} g_{k_r} \Big| +
\Big| h_n g_n +
\sum_s h_{k_s} g_{k_s} \Big| \Big)\;.\nonumber\\
\end{eqnarray}
Any additional splitting of the parties
in the states $i$,
$\hat\rho=\sum_i \eta_i\, \hat\rho_{i,k_r \cdots m}
\otimes \cdots \otimes\hat\rho_{i,k_{r'}}
\otimes\hat\rho_{i,k_s\cdots n}
\otimes \cdots \otimes\hat\rho_{i,k_{s'}}$,
would in general make the bound larger,
eventually yielding the bound for the
fully separable state,
$\sum_j|h_j g_j|/2$ ($j=1...N$).

As mentioned previously, the well-known
bipartite combinations applied to modes $(m,n)$,
$\hat x_m - \hat x_n$ and $\hat p_m + \hat p_n$, mean
all $h_l=g_l=0$ except $h_m=g_m=1$
and $h_n=-g_n=-1$ in Eq.~(\ref{finalmultiparty})
and hence $\langle(\Delta\hat{u})^2\rangle_{\rho}+
\langle(\Delta\hat{v})^2\rangle_{\rho}\geq 1$.

As for a simple example, we may extend that
from the previous section to $N$ modes and set
$h_1=g_1=1$ and $g_2=g_3=\cdots =g_N=
-h_2=-h_3=\cdots =-h_N=1/\sqrt{N-1}$.
Without loss of generality, we choose $m=1$
and obtain for a state of the form 
Eq.~(\ref{Npartyassumption}),
\begin{eqnarray}\label{Npartyexampleonecond}
&&\langle(\Delta\hat{u})^2\rangle_{\rho}+
\langle(\Delta\hat{v})^2\rangle_{\rho}
\nonumber\\
&&\quad\geq\frac{1}{2}\,
\Big(\Big| 1 - \frac{M_r}{N-1} \Big| +
\Big| \frac{1 + M_s}{N-1}\Big| \Big)\;,\nonumber\\
\end{eqnarray}
where $M_r$ is the number of modes potentially
entangled with mode $m=1$ in the convex sum
and $M_s$ is the number of
modes potentially entangled with mode $n$
in the convex sum.
Apart from the fully inseparable case $M_r=N-1$,
the boundary in Eq.~(\ref{Npartyexampleonecond})
is always greater than zero allowing for an
ultimate nonzero bound for all kinds of partial 
separability. Since $[\hat u, \hat v]=0$,
genuine $N$-party entanglement can be verified
when $\langle(\Delta\hat{u})^2\rangle_{\rho}+
\langle(\Delta\hat{v})^2\rangle_{\rho}$ is
sufficiently close to zero.
The ultimate (smallest) bound is given by the
state with the maximum number of modes $M_r$
inseparable from mode $m=1$ in the convex sum,
$M_r=N-2$, and hence $M_s=0$.
This bound is then $1/(N-1)$.  
If none of the modes is inseparable from mode $m=1$,
$M_r=0$ and $M_s=N-2$, the boundary becomes simply that
of a fully separable state, namely one.
Thus, again the violation of a {\it single} condition,
\begin{eqnarray}\label{Npartyexample}
&&\langle\{\Delta[\hat x_1 - (\hat x_2 + \hat x_3 +\cdots 
+\hat x_N)/\sqrt{N-1}]\}^2
\rangle_{\rho}\nonumber\\
&&\;+\langle\{\Delta[\hat p_1 + (\hat p_2 + \hat p_3 +\cdots 
+\hat p_N)/\sqrt{N-1}]\}^2
\rangle_{\rho}\nonumber\\
&&\;\quad\quad\quad\quad\quad\quad\geq 1/(N-1)\;,
\end{eqnarray}
is sufficient for genuine $N$-partite entanglement.
As an example for the violation of the ultimate
bound for genuine $N$-party entanglement, consider 
the $N$-mode state that emerges
after symmetrically splitting one half
of an infinitely squeezed two-mode squeezed state
by $N-2$ beam splitters. The output state is a simultaneous 
eigenstate of $\hat u$ and $\hat v$. In this case,
the EPR correlations, $\hat x_1 - \hat x_2\rightarrow 0$ and
$\hat p_1 + \hat p_2\rightarrow 0$, are transformed into
the $N$-mode correlations
$\hat x_1 - (\hat x_2' + \hat x_3' +\cdots 
+\hat x_N')/\sqrt{N-1}\rightarrow 0$ and
$\hat p_1 + (\hat p_2' + \hat p_3' +\cdots 
+\hat p_N')/\sqrt{N-1}\rightarrow 0$.
A more symmetric example is where both halves
of an EPR state are symmetrically split at beam splitters.
The appropriate combinations are then for instance for four modes,
$\hat x_1' + \hat x_2' - \hat x_3' - \hat x_4'$ and
$\hat p_1' + \hat p_2' + \hat p_3' + \hat p_4'$, also having zero
commutator $[\hat u, \hat v]=0$.
As a further example, we will now discuss
the cv GHZ-type states with quadrature correlations
analogous to those of dv GHZ states.

\section{Example: cv GHZ-type states}

\begin{figure}[tb]
\epsfxsize=1.0in
\epsfbox[200 20 400 550]{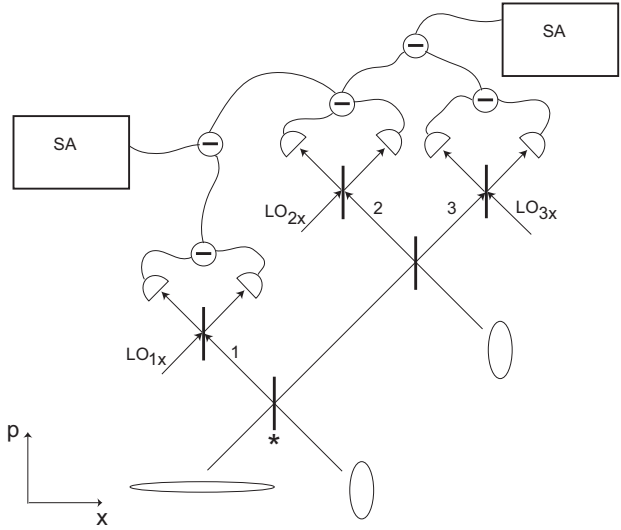}
\caption{\label{fig1}
Verification of genuine tripartite cv entanglement.
$x$ measurements: directly detecting the $x$ quadratures
of all three modes and electronically combining them
in an appropriate way. The three-mode tripartite entangled state of
modes 1, 2, and 3 in this figure is produced with three squeezers and
two beam splitters (the star denotes a $1:2$ BS).}
\end{figure}

\begin{figure}[tb]
\epsfxsize=1.0in
\epsfbox[200 20 400 550]{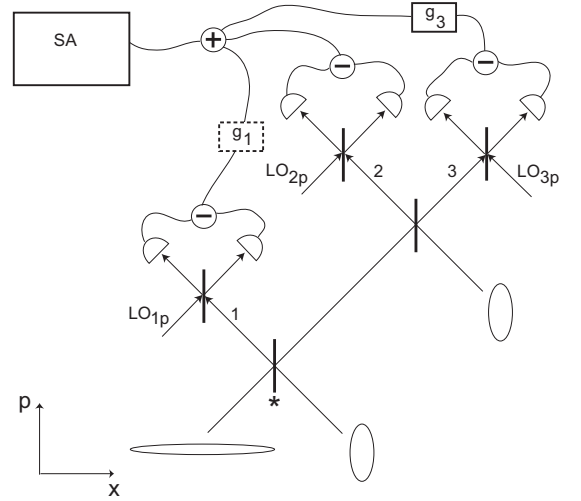}
\caption{\label{fig2}
Verification of genuine tripartite cv entanglement.
$p$ measurements: directly detecting
the $p$ quadratures of all three modes and electronically combining
them in an appropriate way. The three-mode tripartite entangled state of
modes 1, 2, and 3 in this figure is produced with three squeezers and
two beam splitters. The parameters $g_i$ are the ``gains''
from the conditions in Eq.~(\ref{3partycritGHZ}) which can be chosen
optimally (see the text later).}
\end{figure}

We consider a family of genuinely $N$-party entangled states.
The members of this family are those states that emerge from
a particular sequence of $N-1$
phase-free beam splitters (``$N$-splitter'')
with $N$ squeezed state inputs \cite{PvLPRL00}.
By choosing the squeezing direction of one distinct input mode
orthogonal to that of the remaining input modes
(mode 1 squeezed in $p$ and the other modes squeezed in $x$,
as shown in Figs.~\ref{fig1} and \ref{fig2} for $N=3$)
and the degree of squeezing $r_1$ of mode 1
potentially different from that of the other modes
(which are equally squeezed by $r_2$) \cite{Bowen01},
the output states have the following properties
\cite{vanloockFdP02,vanloockcvbook02}.
They are pure $N$-mode states,
totally symmetric under interchange of modes,
and retain the Gaussian character of the input states.
Hence they are entirely described by their second-moment
correlation matrix,
\begin{eqnarray}\label{GHZcorr}
V^{(N)}=\frac{1}{4}
\left( \begin{array}{ccccccc}
a & 0 & c & 0 & c & 0
& \cdots \\
0 & b & 0 & d & 0 & d
& \cdots \\
c & 0 & a & 0 & c & 0
& \cdots \\
0 & d & 0 & b & 0 & d
& \cdots \\
c & 0 & c & 0 & a & 0
& \cdots \\
0 & d & 0 & d & 0 & b
& \cdots \\
\vdots & \vdots & \vdots & \vdots & \vdots & \vdots & \vdots \\
\end{array} \right)\;,
\end{eqnarray}
where
\begin{eqnarray}\label{GHZcorrelements}
a&=&\frac{1}{N}e^{+2r_1}+\frac{N-1}{N}e^{-2r_2}\;,\nonumber\\
b&=&\frac{1}{N}e^{-2r_1}+\frac{N-1}{N}e^{+2r_2}\;,\nonumber\\
c&=&\frac{1}{N} (e^{+2r_1} - e^{-2r_2})\;,\nonumber\\
d&=&\frac{1}{N} (e^{-2r_1} - e^{+2r_2})\;.
\end{eqnarray}
For squeezed vacuum inputs, the multi-mode output states
have zero mean and their Wigner function is of the form
Eq.~(\ref{Gausswigndef}). The particularly simple form
of the correlation matrix in
Eq.~(\ref{GHZcorr}) is, in addition to the
general correlation matrix properties,
symmetric with respect to all modes and contains no
intermode or intramode $x$-$p$ correlations
(hence only four parameters $a$, $b$, $c$, and $d$ are
needed to determine the matrix). However,
the states of this form are in general biased with respect
to $x$ and $p$ ($a\neq b$).
Only for a particular relation between the squeezing
values $(r_1,r_2)$ \cite{vanloockFdP02,vanloockcvbook02},
\begin{eqnarray}\label{Bowenrelation}
e^{\pm 2 r_1}&=&(N-1)\\
&&\times \sinh 2r_2 \,\left[
\sqrt{1+\frac{1}{(N-1)^2\sinh^2 2r_2}} \pm 1 \right],
\nonumber
\end{eqnarray}
the states are unbiased
(all diagonal entries of the correlation matrix equal),
thus having minimum energy at a given degree of entanglement
or, in other words, maximum entanglement for a given mean photon number
\cite{Bowen01}. The other $N$-mode states
of the family can be converted into the minimum-energy state via
local squeezing operations \cite{Bowen01,vanloockFdP02,vanloockcvbook02}.
Only for $N=2$, we obtain $r=r_1=r_2$. In that case,
the matrix $V^{(N)}$ reduces to that of a two-mode squeezed state
which is the maximally entangled state of two modes at
a given mean energy with the correlation matrix entries
$a=b=\cosh 2r$ and $c=\sinh 2r=-d$.
For general $N$, the first squeezer with $r_1$ and the $N-1$
remaining squeezers with $r_2$ have different
squeezing. In the limit of large squeezing
($\sinh 2r_2\approx e^{+2r_2}/2$), we obtain approximately
\cite{vanloockFdP02,vanloockcvbook02}
\begin{eqnarray}\label{Bowenrelationlargesq}
e^{+ 2 r_1}\approx (N-1) e^{+ 2 r_2}  \;.
\end{eqnarray}
For the whole family of $N$-party $N$-mode states with the
correlation matrix in Eq.~(\ref{GHZcorr}),
the quadrature combinations relevant
for detecting genuine multi-party entanglement
are
\cite{PvLPRL00,vanloockFdP02,vanloockcvbook02}
\begin{eqnarray}\label{corr3}
&&\quad\quad\quad\quad\quad\quad\quad\quad\quad\;
\langle[\Delta(\hat{x}_m-\hat{x}_n)]^2\rangle=e^{-2r_2}/2\;,
\nonumber\\
&&\left\langle\left[\Delta
\left(\hat{p}_m+\hat{p}_n+g^{(N)}\sum^N_{j\neq m,n}
\hat{p}_j\right)\right]^2\right\rangle=\nonumber\\
&&\frac{[2+(N-2)g^{(N)}]^2}{4N}e^{-2r_1}
+\frac{(g^{(N)}-1)^2(N-2)}{2N}e^{+2r_2}\;.\nonumber\\
\end{eqnarray}
The total variances are then optimized (minimized) for
\begin{eqnarray}\label{optgain}
g_{\rm opt}^{(N)}=
\frac{e^{+2r_2}-e^{-2r_1}}{e^{+2r_2}+\frac{N-2}{2}\,e^{-2r_1}}
\;.
\end{eqnarray}
In the limit of infinite squeezing, $r_1,r_2\to\infty$, the
above correlations correspond to a simultaneous
eigenstate of the relative positions and the total momentum
such as the cv GHZ states in Eq.~(\ref{PVLmaxentGHZbasis}).

Let us now examine how to experimentally verify the genuine multipartite
entanglement of the cv GHZ-type states
(in any case, it may be verified in an operational way by doing quantum
teleportation between every pair of parties with the help of the remaining
party \cite{PvLPRL00}). Due to experimental imperfections,
we may assume that the entanglement of slightly degraded approximate versions
of the states generated according to a scheme as in Figs.~\ref{fig1}
and \ref{fig2} is to be verified.
We start again with only three parties and modes.
For a simple check, look at the following set of
inequalities,
\begin{eqnarray}\label{3partycritGHZ}
{\rm I.}\;\;\quad
\langle[\Delta(\hat{x}_1-\hat{x}_2)]^2\rangle +
\langle[\Delta(\hat{p}_1+\hat{p}_2+g_3\hat{p}_3)]^2\rangle\geq 1\;,
\nonumber\\
{\rm II.}\;\quad
\langle[\Delta(\hat{x}_2-\hat{x}_3)]^2\rangle +
\langle[\Delta(g_1\hat{p}_1+\hat{p}_2+\hat{p}_3)]^2\rangle\geq 1\;,
\nonumber\\
{\rm III.}\quad
\langle[\Delta(\hat{x}_1-\hat{x}_3)]^2\rangle +
\langle[\Delta(\hat{p}_1+g_2\hat{p}_2+\hat{p}_3)]^2\rangle\geq 1\;.
\nonumber\\
\end{eqnarray}
On the l.h.s. of condition I., 
we have $h_1=-h_2=g_1=g_2=1$ and $h_3=0$,
and hence the boundary for the total variance
in Eq.~(\ref{3partycritgenansatz})
becomes one with Eq.~(\ref{statem2}) and Eq.~(\ref{statem3}),
but zero with Eq.~(\ref{statem1}).
Similarly, using the l.h.s. of condition II., where
$h_2=-h_3=g_2=g_3=1$ and $h_1=0$,
the boundary is one for Eq.~(\ref{statem1}) and 
Eq.~(\ref{statem2}), but zero for Eq.~(\ref{statem3}).
Finally, the l.h.s of condition III. 
with $h_1=-h_3=g_1=g_3=1$ and $h_2=0$
corresponds to a boundary of one 
in Eq.~(\ref{statem1}) and Eq.~(\ref{statem3}),
and a boundary of zero in Eq.~(\ref{statem2}).
Thus, the following statements
for (at least partially) separable states hold,
\begin{eqnarray}\label{3partystatementsGHZ}
\hat\rho&=&\sum_i \eta_i\, \hat\rho_{i,12}\otimes\hat\rho_{i,3}
\quad \rightarrow {\rm II.}\; {\rm and}\; {\rm III.}\;,
\nonumber\\
\hat\rho&=&\sum_i \eta_i\, \hat\rho_{i,13}\otimes\hat\rho_{i,2}
\quad \rightarrow {\rm I.}\; {\rm and}\; {\rm II.}\;,
\nonumber\\
\hat\rho&=&\sum_i \eta_i\, \hat\rho_{i,23}\otimes\hat\rho_{i,1}
\quad \rightarrow {\rm I.}\; {\rm and}\; {\rm III.}
\end{eqnarray}
The conditions in Eq.~(\ref{3partycritGHZ}) are necessary for
different kinds of partial separability. As a result,
the violation of {\it any pair} of inequalities in 
Eq.~(\ref{3partycritGHZ}) is sufficient for genuine three-party
three-mode entanglement.
Violating only one condition in Eq.~(\ref{3partycritGHZ}) 
(for example, condition I.) means that the total density
operator cannot be written in two of the three forms
in Eq.~(\ref{3partystatementsGHZ}) 
(for example, neither in the form
$\hat\rho=\sum_i \eta_i\, \hat\rho_{i,13}\otimes\hat\rho_{i,2}$
nor in the form
$\hat\rho=\sum_i \eta_i\, \hat\rho_{i,23}\otimes\hat\rho_{i,1}$).
Using the classification of Ref.~\cite{Giedke01}, 
the classes 3 [two-mode biseparable states expressible
in two of the three forms in Eq.~(\ref{3partystatementsGHZ})],
4 [three-mode biseparable states expressible
in all of the three forms in Eq.~(\ref{3partystatementsGHZ})], 
and 5 [fully separable states describable by
Eq.~(\ref{3partyfullysep})] are then ruled out.
The forms of the classes 1 (fully inseparable states) and 2 
[one-mode biseparable states expressible
in one of the three forms in Eq.~(\ref{3partystatementsGHZ})]
remain both possible. In our example with the violation of I.,
the state might be genuinely tripartite entangled or
of the partially separable form 
$\hat\rho=\sum_i \eta_i\, \hat\rho_{i,12}\otimes\hat\rho_{i,3}$.
Eventually, the violation of a second inequality   
in Eq.~(\ref{3partycritGHZ}) (for instance, condition II.)
negates also the only remaining partially separable form (e.g., 
$\hat\rho=\sum_i \eta_i\, \hat\rho_{i,12}\otimes\hat\rho_{i,3}$),
thus proving the full inseparability of the state
\cite{noteonchineseexp}. 
Note that even though pure and totally 
symmetric multi-party entangled states are always genuinely multipartite
entangled \cite{vanloockFdP02,vanloockcvbook02}, asymmetric pure
or mixed entangled three-mode states
(e.g., from class 2 in Ref.~\cite{Giedke01}, the product state of a 
bipartite entangled two-mode squeezed state and a vacuum state) and
symmetric mixed entangled three-mode states 
(like the example for the three-mode 
biseparable class, class 4, given in Ref.~\cite{Giedke01})
do not automatically exhibit genuine tripartite entanglement.
Due to the violation of {\it two} conditions in 
Eq.~(\ref{3partycritGHZ}),
the two loopholes of partial separability, mixedness
and/or asymmetry, are ruled out.

The criteria here are only sufficient for full 
inseparability and hence genuinely tripartite entangled states
may also satisfy all the conditions in Eq.~(\ref{3partycritGHZ})
(an example will be mentioned later). On the other hand,
note that we did not use the assumption of Gaussian states. 
The derivation of the conditions relies only on the Cauchy-Schwarz
inequality and Heisenberg's (sum) uncertainty relation.  

Alternatively, one could simply check the known
bipartite separability conditions \cite{Duan00} 
for pairs of modes, i.e., $g_1=g_2=g_3=0$
in Eq.~(\ref{3partycritGHZ})
(or using products of variances \cite{Tan99} instead of sums).
Again, the statements in Eq.~(\ref{3partystatementsGHZ}) hold.
Hence two violations again verify genuine tripartite entanglement.
However, the significance of the more general conditions
in Eq.~(\ref{3partycritGHZ}) compared to those with
$g_1=g_2=g_3=0$ is that for the cv GHZ-type states,
as discussed later, 
the former can be {\it always} violated
for any degree of multi-party entanglement
and the violations can steadily grow from 
small towards ``perfect'' (that is all variances
of the combinations zero) as the three-mode entanglement  
increases. In contrast, the bipartite conditions
with $g_1=g_2=g_3=0$ may be violated
for bad three-mode entanglement (small squeezing)
and satisfied for larger squeezing, thus not always
verifying genuine tripartite entanglement, 
and in particular never verifying
good genuine tripartite entanglement.
Moreover, they might be always violated, but the violations
do not attain a significant amount 
(e.g., three-mode states made from one squeezed state
\cite{vanloockFdP02,vanloockcvbook02}).
Similarly, using products of variances \cite{Tan99} instead of
sums in Eq.~(\ref{3partycritGHZ})
with $g_1=g_2=g_3=0$, violations may always
occur, but also only to a certain extent 
\cite{vanloockFdP02,vanloockcvbook02}.
In Figs.~\ref{fig1} and \ref{fig2}, it is shown how to apply
the tripartite entanglement criteria experimentally
using homodyne detectors.

Let us also discuss the conditions for the $N=4$ case
in more detail.
We consider a set of six inequalities,
\begin{eqnarray}\label{4partycrit}
{\rm I.}\;\;\,
\langle[\Delta(\hat{x}_1-\hat{x}_2)]^2\rangle +
\langle[\Delta(\hat{p}_1+\hat{p}_2+g_3\hat{p}_3
+g_4\hat{p}_4)]^2\rangle\geq 1,
\nonumber\\
{\rm II.}\;\,
\langle[\Delta(\hat{x}_2-\hat{x}_3)]^2\rangle +
\langle[\Delta(g_1\hat{p}_1+\hat{p}_2+\hat{p}_3
+g_4\hat{p}_4)]^2\rangle\geq 1,
\nonumber\\
{\rm III.}\,
\langle[\Delta(\hat{x}_1-\hat{x}_3)]^2\rangle +
\langle[\Delta(\hat{p}_1+g_2\hat{p}_2+\hat{p}_3
+g_4\hat{p}_4)]^2\rangle\geq 1,
\nonumber\\
{\rm IV.}\,
\langle[\Delta(\hat{x}_3-\hat{x}_4)]^2\rangle +
\langle[\Delta(g_1\hat{p}_1+g_2\hat{p}_2+\hat{p}_3
+\hat{p}_4)]^2\rangle\geq 1,
\nonumber\\
{\rm V.}\;
\langle[\Delta(\hat{x}_2-\hat{x}_4)]^2\rangle +
\langle[\Delta(g_1\hat{p}_1+\hat{p}_2+g_3\hat{p}_3
+\hat{p}_4)]^2\rangle\geq 1,
\nonumber\\
{\rm VI.}\;
\langle[\Delta(\hat{x}_1-\hat{x}_4)]^2\rangle +
\langle[\Delta(\hat{p}_1+g_2\hat{p}_2+g_3\hat{p}_3
+\hat{p}_4)]^2\rangle\geq 1.
\nonumber\\
\end{eqnarray}
The position and momentum variables $\hat x_l$ and $\hat p_l$
are the quadratures of four electromagnetic modes this time.
The $g_l$ are again arbitrary real parameters.
Now the following statements
for (at least partially) separable states hold,
\begin{eqnarray}\label{4partystatements1}
\hat\rho&=&\sum_i \eta_i\, \hat\rho_{i,123}\otimes\hat\rho_{i,4}
\;\rightarrow {\rm IV.,}{\rm V.,}\;
{\rm and}\; {\rm VI.,}\;\nonumber\\
\hat\rho&=&\sum_i \eta_i\, \hat\rho_{i,124}\otimes\hat\rho_{i,3}
\;\rightarrow {\rm II.,}{\rm III.,}\;
{\rm and}\; {\rm IV.,}\;\nonumber\\
\hat\rho&=&\sum_i \eta_i\, \hat\rho_{i,134}\otimes\hat\rho_{i,2}
\quad\,\rightarrow {\rm I.,}{\rm II.,}\;
{\rm and}\; {\rm V.,}\;\nonumber\\
\hat\rho&=&\sum_i \eta_i\, \hat\rho_{i,234}\otimes\hat\rho_{i,1}
\;\,\rightarrow {\rm I.,}{\rm III.,}\;
{\rm and}\; {\rm VI.,}\;\nonumber\\
\end{eqnarray}
and,
\begin{eqnarray}\label{4partystatements2}
\hat\rho&=&\sum_i \eta_i\, \hat\rho_{i,12}\otimes\hat\rho_{i,34}
\;\rightarrow {\rm II.,}{\rm III.,}{\rm V.,}\;
{\rm and}\; {\rm VI.,}\;\nonumber\\
\hat\rho&=&\sum_i \eta_i\, \hat\rho_{i,13}\otimes\hat\rho_{i,24}
\quad\rightarrow {\rm I.,}{\rm II.,}{\rm IV.,}\;
{\rm and}\; {\rm VI.,}\;\nonumber\\
\hat\rho&=&\sum_i \eta_i\, \hat\rho_{i,14}\otimes\hat\rho_{i,23}
\quad\,\rightarrow {\rm I.,}{\rm III.,}{\rm IV.,}\;
{\rm and}\; {\rm V.,}\;\nonumber\\
\end{eqnarray}
and finally,
\begin{eqnarray}\label{4partystatements3}
\hat\rho&=&\sum_i \eta_i\, \hat\rho_{i,12}\otimes\hat\rho_{i,3}
\otimes\hat\rho_{i,4}
\;\rightarrow {\rm all}\;{\rm except}\;{\rm I.,}
\nonumber\\
\hat\rho&=&\sum_i \eta_i\, \hat\rho_{i,13}\otimes\hat\rho_{i,2}
\otimes\hat\rho_{i,4}
\;\rightarrow {\rm all}\;{\rm except}\;{\rm III.,}
\nonumber\\
\hat\rho&=&\sum_i \eta_i\, \hat\rho_{i,14}\otimes\hat\rho_{i,2}
\otimes\hat\rho_{i,3}
\;\rightarrow {\rm all}\;{\rm except}\;{\rm VI.,}
\nonumber\\
\hat\rho&=&\sum_i \eta_i\, \hat\rho_{i,23}\otimes\hat\rho_{i,1}
\otimes\hat\rho_{i,4}
\;\rightarrow {\rm all}\;{\rm except}\;{\rm II.,}
\nonumber\\
\hat\rho&=&\sum_i \eta_i\, \hat\rho_{i,24}\otimes\hat\rho_{i,1}
\otimes\hat\rho_{i,3}
\;\rightarrow {\rm all}\;{\rm except}\;{\rm V.,}
\nonumber\\
\hat\rho&=&\sum_i \eta_i\, \hat\rho_{i,34}\otimes\hat\rho_{i,1}
\otimes\hat\rho_{i,2}
\;\rightarrow {\rm all}\;{\rm except}\;{\rm IV.}
\nonumber\\
\end{eqnarray}
Note that again the fully separable state,
$\hat\rho=\sum_i \eta_i\, \hat\rho_{i,1}\otimes\hat\rho_{i,2}
\otimes\hat\rho_{i,3}\otimes\hat\rho_{i,4}$,
is included.
The above statements can be easily confirmed using
Eq.~(\ref{finalmultiparty}) for states of the general form
Eq.~(\ref{Npartyassumption}).
The different forms here are
$\hat\rho=\sum_i \eta_i\, \hat\rho_{i,klm}\otimes\hat\rho_{i,n}$,
including
$\hat\rho=\sum_i \eta_i\, \hat\rho_{i,kl}\otimes\hat\rho_{i,m}
\otimes\hat\rho_{i,n}$, and
$\hat\rho=\sum_i \eta_i\, \hat\rho_{i,km}\otimes\hat\rho_{i,ln}$,
including
$\hat\rho=\sum_i \eta_i\, \hat\rho_{i,km}\otimes\hat\rho_{i,l}
\otimes\hat\rho_{i,n}$,
with the two modes $m$ and $n$ always being separable.
For the combinations
$\hat u=\hat x_m - \hat x_n$ and
$\hat v= g_k\hat p_k + g_l\hat p_l + \hat p_m 
+ \hat p_n$, the boundary of the total variance is
always one. The statements 
Eq.~(\ref{4partystatements1}), Eq.~(\ref{4partystatements2}),
and Eq.~(\ref{4partystatements3}) become obvious then
by considering all possible pairs of modes $(m,n)$ of the 
four modes $(k,l,m,n)$. Note that always when the two
modes $(m,n)$ are potentially entangled, the boundary for
the total variance drops to zero.

What kind of violations of the six inequalities
in Eq.~(\ref{4partycrit}) are now sufficient
to verify the full inseparability of a
four-mode four-party state?
The violations must rule out any of the 
partially separable forms in 
Eq.~(\ref{4partystatements1}), 
Eq.~(\ref{4partystatements2}), and 
Eq.~(\ref{4partystatements3}).
Let us, for example, consider violations of
the inequalities IV. and V.
These violations mean that all partially separable 
forms in 
Eq.~(\ref{4partystatements1}), 
Eq.~(\ref{4partystatements2}), and 
Eq.~(\ref{4partystatements3}) are excluded except
for the form
$\hat\rho=\sum_i \eta_i\, \hat\rho_{i,234}
\otimes\hat\rho_{i,1}$ in Eq.~(\ref{4partystatements1}).
In order to negate this form as well
a further violation is needed.
According to Eq.~(\ref{4partystatements1}),
one of the inequalities I., III., or VI.
should be violated in addition.
Here it is important to realize that 
the conditions IV. and V. do not involve the $x$ quadrature
of mode 1, but that of all the other modes.
The additional test via any one of the conditions
I., III., or VI., of which all contain 
both quadratures of mode 1, eventually 
provides the missing information about mode 1. Hence 
we learn that three conditions are sufficient here
to verify the full inseparability of a
four-mode four-party state.
We may choose
\begin{eqnarray}\label{4partycritfinal}
\langle[\Delta(\hat{x}_1-\hat{x}_2)]^2\rangle +
\langle[\Delta(\hat{p}_1+\hat{p}_2+g_3\hat{p}_3
+g_4\hat{p}_4)]^2\rangle < 1,
\nonumber\\
\langle[\Delta(\hat{x}_2-\hat{x}_3)]^2\rangle +
\langle[\Delta(g_1\hat{p}_1+\hat{p}_2+\hat{p}_3
+g_4\hat{p}_4)]^2\rangle < 1,
\nonumber\\
\langle[\Delta(\hat{x}_3-\hat{x}_4)]^2\rangle +
\langle[\Delta(g_1\hat{p}_1+g_2\hat{p}_2+\hat{p}_3
+\hat{p}_4)]^2\rangle < 1,
\nonumber\\
\end{eqnarray}
which involve both quadratures $x$ and $p$ of all four modes.
Note that apart from the coefficients $g_l$,
these four combinations correspond to those observables 
measured in a four-party cv GHZ state analyzer.
Correspondingly, for $N$ parties and modes,
we may choose the following $N-1$ conditions
in terms of effectively $N$ combinations
(those of an $N$-party $N$-mode cv GHZ state analyzer),
\begin{eqnarray}\label{Npartycritfinal}
&&\langle[\Delta(\hat{x}_1-\hat{x}_2)]^2\rangle 
\nonumber\\
&&+
\langle[\Delta(\hat{p}_1+\hat{p}_2+g_3\hat{p}_3
+\cdots +g_N\hat{p}_N)]^2\rangle < 1,
\nonumber\\
&&\langle[\Delta(\hat{x}_2-\hat{x}_3)]^2\rangle 
\nonumber\\
&&+
\langle[\Delta(g_1\hat{p}_1+\hat{p}_2+\hat{p}_3
+g_4\hat{p}_4+\cdots +g_N\hat{p}_N)]^2\rangle < 1,
\nonumber\\
&&\quad\vdots\quad\quad\quad\vdots\quad\quad\quad
\vdots\quad\quad\quad
\vdots\quad\quad\quad\vdots\quad\quad\quad
\vdots\quad\quad\quad\vdots
\nonumber\\
&&\langle[\Delta(\hat{x}_{N-1}-\hat{x}_N)]^2\rangle 
\nonumber\\
&&+
\langle[\Delta(g_1\hat{p}_1+g_2\hat{p}_2+\cdots +
g_{N-2}\hat{p}_{N-2}+\hat{p}_{N-1}+\hat p_N)]^2\rangle 
\nonumber\\
&&\quad\quad\quad\quad\quad\quad\quad\quad\quad
\quad\quad\quad\quad
\quad\quad\quad\quad< 1.
\end{eqnarray}
These conditions are sufficient to verify the
full inseparability (genuine $N$-party entanglement)
of an $N$-party $N$-mode state.
For arbitrary $N$, the proof relies on the fact that
in any partially separable form we may always select 
a distinct pair of modes $(m,n)$ which are
separable in the states $i$ of the convex sum
of the density operator.
Only exploiting that modes $m$ 
and $n$ are separable, the combinations
\begin{eqnarray}\label{Npartycombin}
\hat u=\hat x_m - \hat x_n \;,\quad
\hat v=\sum_{j=1}^{N-2}
g_{k_j}\hat p_{k_j} + \hat p_m + \hat p_n\;,
\end{eqnarray}
always yield a boundary of one for the total variance
using Eq.~(\ref{finalmultiparty}) for states of the general 
form Eq.~(\ref{Npartyassumption}).
By taking the pairs of modes $(1,2)$, $(2,3)$, ...,
$(N-1,N)$ for $(m,n)$, all partially separable forms
of the total density operator are covered 
(as demonstrated explicitly for $N=4$) and hence
the $N-1$ conditions in Eq.~(\ref{Npartycritfinal}) are 
sufficient for genuine $N$-party $N$-mode inseparability.

\begin{figure}[t]
\epsfxsize=3.8in
\epsfbox[80 10 400 190]{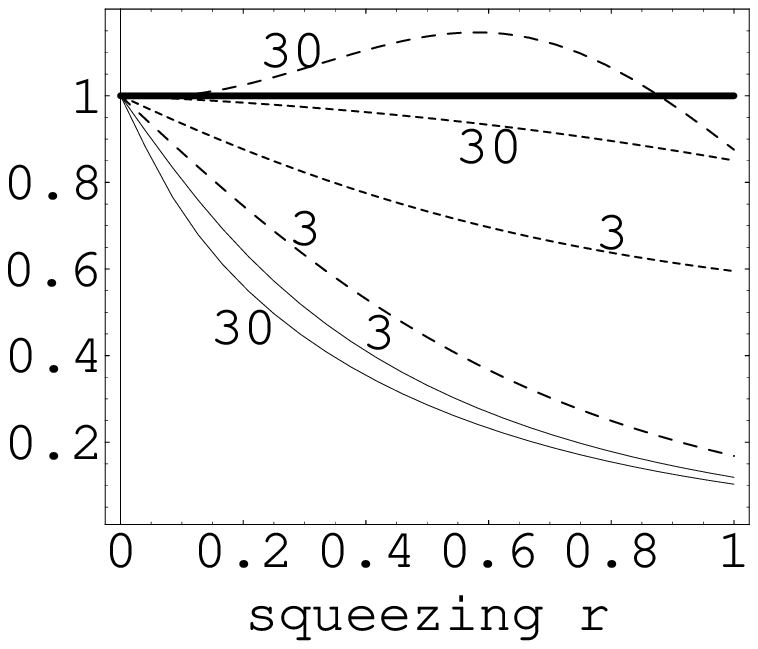}
\caption{\label{fig3}
Plot of the left-hand-side (total variance) of the
conditions in Eq.~(\ref{Npartycritfinal})
for different $N$-mode states
with quadrature correlations given by
Eq.~(\ref{corr3})
and different numbers of
parties $N=3$ and $N=30$. The states are those produced with one
squeezed state (dotted lines with $r_2=0$ and $r=r_1$),
those made from $N$ equally squeezed states
(dashed lines with $r=r_1=r_2$), and the unbiased minimum-energy
states with squeezing $r_1$ and $r=r_2$ related as in
Eq.~(\ref{Bowenrelation}).}
\end{figure}

The left-hand-side of the
inequalities in Eq.~(\ref{Npartycritfinal})
is shown in Fig.\ref{fig3}
for various cv GHZ-type $N$-mode states differing in the
relation between the squeezing $r_1$ and $r_2$
[Eq.~(\ref{GHZcorr}), Eq.~(\ref{GHZcorrelements}), and
Eq.~(\ref{corr3})].
Due to the total symmetry of all these states,
the left-hand-side of the conditions in Eq.~(\ref{Npartycritfinal})
becomes equal for all conditions (assuming $g_j\equiv g^{(N)}$).
Hence values below the boundary 1 here mean all inequalities
in Eq.~(\ref{Npartycritfinal}) are satisfied,
thus indicating genuine $N$-party entanglement.
In all cases in Fig.~\ref{fig3}, the optimal coefficients $g_j\equiv
g_{\rm opt}^{(N)}$ from Eq.~(\ref{optgain})
are used to minimize the total variances of Eq.~(\ref{corr3}).
If $N=30$, only for the unbiased states,
the conditions are always met (for any nonzero squeezing $r>0$) {\it and}
the total variances tend to zero for large squeezing.
Moreover, for the same squeezing $r$, the unbiased states with $N=30$
drop below the boundary 1 to a greater extent
than their unbiased tripartite counterparts.
In contrast, for the biased states (those with only
one squeezer, $r_2=0$ and $r=r_1$, and those with $N$ equally
squeezed states, $r=r_1=r_2$), the total variances approach
or even exceed the boundary 1 as the number of parties grows.
The example of the states with $N$ equal squeezers also demonstrates
that there are Gaussian states which are indeed genuinely
$N$-party entangled, but do not satisfy any of the
conditions in Eq.~(\ref{Npartycritfinal}).
It can be shown, however, taking into account the symmetry and purity
of the whole family of $N$-mode states (including those with $N$ equal
squeezers) that all these states are genuinely multi-party entangled
for any nonzero squeezing \cite{vanloockFdP02,vanloockcvbook02}.

Finally, we emphasize that one may use other conditions too
for verifying the genuine multipartite entanglement
of the cv GHZ-type states.
Even a single condition might be again sufficient.
For example, consider the combinations
$\hat u=2\hat x_1 -(\hat x_2 + \hat x_3)$ and 
$\hat v=\hat p_1 + \hat p_2 + \hat p_3$ for three modes.
We have $[\hat u,\hat v]=0$, and indeed the GHZ-type three-mode
state becomes a simultaneous eigenstate of $\hat u$ and $\hat v$
in the limit of infinite squeezing, $r_1,r_2\to\infty$.
The boundaries of the total variance for these
combinations take on the value one when
$\hat\rho=\sum_i \eta_i\, \hat\rho_{i,12}\otimes\hat\rho_{i,3}$
or
$\hat\rho=\sum_i \eta_i\, \hat\rho_{i,13}\otimes\hat\rho_{i,2}$,
and the value two (corresponding to the fully separable state)
when
$\hat\rho=\sum_i \eta_i\, \hat\rho_{i,23}\otimes\hat\rho_{i,1}$.
Hence $\langle(\Delta\hat{u})^2\rangle_{\rho}+
\langle(\Delta\hat{v})^2\rangle_{\rho}<1$ is sufficient
for genuine tripartite entanglement.
The number of measurements required, however, remain the 
same as for the criteria above expressed by $N-1$ conditions.
In any case, both quadratures of all modes must be detected
and combined in an appropriate way.

\section{Conclusions}

In summary, we proposed experimental criteria to detect
genuine multipartite continuous-variable entanglement.
These are expressed in terms of the variances of
particular combinations of all the quadratures involved.
The combinations are measurable with only a few simple 
homodyne detections. For Gaussian states, it is then not necessary 
to determine the entire correlation matrix in order to confirm the 
genuine multipartite entanglement.
Furthermore, the conditions here do not rely on the
assumption of Gaussian states. An experimental confirmation
of the Gaussian character of the state in question is therefore 
not needed either. Finally, we examined the applicability
of the conditions to a particular GHZ-type class of 
genuinely multi-party entangled states.
These states are of Gaussian form, they are totally symmetric under 
exchange of modes, and they have zero cross correlations between 
the $x$ and the $p$ quadratures.
If they are in addition unbiased between the $x$ and the $p$
quadratures, they always (for any nonzero entanglement)
satisfy the conditions in terms of appropriately chosen linear 
combinations. In the limit of perfect entanglement,
the variances of the combinations tend to zero
for the unbiased states and the conditions are perfectly met.

In an experiment, one normally has approximate a priori
knowledge about the state to be analyzed.
According to this a priori knowledge, one can then choose
appropriate linear combinations to be measured.
It would be desirable to know whether there is always,
for any given multi-party multi-mode state, 
a single optimal condition to verify its genuine multipartite 
entanglement and how to constructively derive this condition.
Inferring from the results here, such a condition may always
exist and the corresponding linear combinations must contain
both quadratures of all modes with optimized coefficients
$h_l$ and $g_l$.
A possible approach to this question is in terms
of so-called entanglement witnesses 
\cite{Horodecki96,Terhal00}. One may then
interpret the inequalities for the total variances
as quantum expectation values of Hermitian operators
which take on negative values when they witness some kind
of partial inseparability.

{\it Acknowledgements:} 
PvL is grateful to Masahide Sasaki, Masahiro Takeoka,
Marcos Curty, Norbert L\"{u}tkenhaus, and Samuel Braunstein for
useful discussions.
AF acknowledges the financial support of 
MPHPT and MEXT of Japan. PvL thanks the Communications Research 
Laboratory Tokyo for funding a research visit.
He also acknowledges the financial support of
the DFG under the Emmy-Noether programme.

\end{document}